\documentclass[submission, LectureNotes]{SciPost}

\usepackage{amsmath}  
\usepackage{amsfonts} 
\usepackage{graphicx} 

\newcommand{\ist}[1]{\overset{\footnotesize(\ref{#1})}{=}}
\newcommand{\iist}[2]{\overset{^{(\ref{#1})}}{\underset{^{(\ref{#2})}}{=}}}

\begin{document}

\begin{center}{\Large \textbf{
Dirac monopoles and the importance of the usage of appropriate degrees of freedom
}}\end{center}

\begin{center}
M. Faber\textsuperscript{1*},
M. Suda\textsuperscript{2}
\end{center}

\begin{center}
{\bf 1} Atominstitut, Techn. Univ. Wien, Operngasse 9, A-1040 Wien, Austria
\\
{\bf 2} AIT Austrian Institute of Technology GmbH, Digital Safety \&
Security, Security \& Communication Technologies,
Giefinggasse 4, 1210 Vienna, Austria
\\
* faber@kph.tuwien.ac.at
\end{center}

\begin{center}
\today
\end{center}


\section*{Abstract}
{\bf
We discuss that the singularities appearing in Dirac's formulation of magnetic monopoles are due to the set of fields which he used and not due to the physical properties of magnetic monopoles. We explain in detail that we can find the same algebraic expressions and singularities for the affine connections on the sphere $S^2$, which Dirac found for the U(1) gauge field of magnetic monopoles. Since spheres have no singularities, it is obvious, that these singularities are due to the set of fields which are used to describe the geometry of $S^2$. As there are descriptions of the geometry of spheres without any singularities, we indicate that it would be preferable to use singularity free descriptions of magnetic and also electric monopoles.
}

\vspace{10pt}
\noindent\rule{\textwidth}{1pt}
\tableofcontents\thispagestyle{fancy}
\noindent\rule{\textwidth}{1pt}
\vspace{10pt}

\section{Introduction}\label{SecEinf}
In several papers~\cite{Faber:2017uvr,Faber:2014bxa,Faber:2008hr,Borisyuk:2007bd,Faber:2002nw,Faber:1999ia} the authors have shown that the three rotational degrees of freedom (DOFs) of a spatial Dreibein (triad) in 3+1 dimensional space-time are sufficient to get an interesting model for electromagnetic phenomena. The double covering group of the rotational group SO(3) is SU(2)$~\cong~S^3$. The three DOFs of SO(3) resp. SU(2) allow for three topological quantum numbers of fields in space-time. The non-trivial mappings to rotational degrees of freedom are related to the homotopy groups
\begin{itemize}
\item $\pi_2(S^2)=\mathbb Z~\to~$charge number
\item $\pi_3(S^2)=\mathbb Z~\to~$photon number
\item $\pi_3(S^3)=\mathbb Z~\to~$spin quantum numbers
\end{itemize}
These three mappings are characterised by topological quantum numbers. With the three DOFs of SO(3) and an appropriate Lagrangian in a dual formulation one can describe topological solitons of finite self energy (mass) with quantized charges and half-integer spins and their field without any singularities and in addition the photon number as a topological quantum number. Due to the experience with this special choice of DOFs the authors want to underline the importance of the choice of appropriate field variables by the discussion of a further example, the description of Dirac monopoles and the comparison to the description of the geometry of $S^2$.

Restricting rotations to angles of $\pi$, the rotational DOFs reduce from $S^3$ to $S^2$, i.e. to normalized three component vector fields on $S^2$, one is loosing the spin quantum number, preserving photon and charge numbers. Charge quantization is well-known from Dirac monopoles. The magnetic charge is quantised without the presence of any electric charge. There exist three descriptions of Dirac monopoles. First: the original description of Dirac~\cite{dirac:1931kp,dirac:1948um} by a U(1) gauge field has two types of singularities, the line-like singularities of Dirac strings and the point-like singularities in the centre of the Dirac monopoles. Second: the fibre-bundle description by Wu and Yang~\cite{wu:1975es,Wu:1976qk} with at least two different gauge fields~\cite{Ghosh:2002wa,Chatterjee:2003xu} enlarges the U(1) DOFs to the DOFs of a fibre bundle, where the fibres are the unphysical gauge DOFs. Third: the description with a normalised three component vector field $\vec n\in S^2$ avoids with only two DOFs the singularities of the Dirac string~\cite{wu:1969wy,Wu:1975vq}. Of the above mentioned topological quantum numbers from $\pi_3(S^2)$ and $\pi_2(S^2)$ usually only the last one, the magnetic charge, is discussed. But as shown in Ref.~\cite{Faber:2017uvr} the $\vec n$-field allows also to describe photons.

``The use of coordinates introduced by Descartes was a great step in mathematics since it allowed to translate geometrical notions into analysis and algebra''~\cite{Zanelli:2005sa}. We know from our experience that it may be very helpful to adjust the choice of the coordinates to the problem, one wants to treat. For spherical symmetric problems it is very helpful to use spherical coordinates. We are familiar with the singularities which appear for azimuthal angles $0$ and $\pi$ and we know that these singularities are coordinate singularities and not geometrical singularities.

In this paper we would like to underline that we find an analogous situation in the mathematical formulation, which Dirac~\cite{dirac:1931kp,dirac:1948um} has chosen to describe magnetic monopoles. We try to demonstrate in a pedagogical approach, that the singularities, which appear in this description, are related to the chosen set of fields and not to the physics of monopoles. We conclude, that for the treatment of every physical problem it is important to choose an appropriate set of fields.

Magnetic monopoles by themselves are very interesting topological objects, they are characterised by a topological quantum number, which corresponds to the magnetic charge $g$. A field with monopoles can only have an integer number of monopoles and antimonopoles. There is no fractional monopole number possible. This property of the quantization of charge is not shared by electric charges, when they are described by Maxwell's electrodynamics. Classical Maxwell's equations allow for arbitrary values of electric charges, whereas we know from Millikan's experiment that the electric charges in nature appear only in integer multiples of the elementary charge $e_0=1.602\times10^{-19}$~C.

Dirac's prediction of the existence of positrons and their experimental discovery reshaped completely our understanding of nature. This success may have inspired him, to formulate a theory of magnetic monopoles symmetrizing Maxwell's equations in the presence of sources. As we describe in Sect.~\ref{SecMonop}, Dirac introduced an unmeasurable singular line, the famous Dirac string, to allow for a formulation of the magnetic field $\vec B$ as curl of the $\vec A$-field. As shown in Sect.~1.3 of Ref.~\cite{Shnir:2005joa} this string corresponds to an semi-infinite and infinitely thin solenoid which transports the magnetic flux to the center of the monopole. If this flux is quantized to the magnetic flux quantum $g=h/e_0$, where $h$ is Planck's constant, this string is not detectable. For other values of magnetic flux, the solenoid would give rise to an Aharonov-Bohm effect, see Ref.~\cite{Aharonov:1959fk}.

Dirac himself wrote in 1981: ``I am inclined now to believe that monopoles do not exist. Too many years have gone without any encouragement from the experimental side.''~\cite{Craigie1982:monopoles} Nevertheless, Dirac monopoles are intensively discussed in several books~\cite{Polyakov:1987ez,Manton:2004tk,Shnir:2005joa} and in a huge number of articles~\cite{Rindler:1989pf,Goldhaber:1989pj,Crawford:1992zs,Singleton:1996ru,deLemos:2009bn,Rajantie:2012,Rajantie:2016paj}. In Quantum Chromo Dynamics they are very important in the discussions about the mechanism for the confinement of color charges. In the dual super conductor model~\cite{Ripka:2003vv} supercurrents of magnetic monopoles confine the gluon string between pairs of color sources.

In Section~\ref{SecMonop} we repeat the well-known formulation of Dirac monopoles by a U(1)-gauge field~\cite{dirac:1931kp,dirac:1948um}. In Section ~\ref{SecS2} we formulate the geometry of $S^2$ with a U(1)-valued affine connection~\cite{Visconti:2003aa}. Finally, in Sect.~\ref{SecSchluss} we mention, which fields should be more appropriate for the description of monopoles and two-dimensional spheres.

\section{Dirac monopoles}\label{SecMonop}
The symmetry of the sourceless Maxwell equations under dual transformations, the exchange of electric $\vec E$ and magnetic fields $\vec B$,  $\vec E\to c\vec B$ and $c\vec B\to-\vec E$, was extended by Dirac to the full Maxwell equations introducing a magnetic charge density $\rho_M$ and magnetic currents $\vec j_M$. Dirac observed the interesting property of monopoles, that their charges are quantized in units of some magnetic charge $g$. He further emphasised the interesting fact, that the product of both charge units is an integer multiple of Planck's constant $h$
\begin{equation}\label{DiracQuantBed}
ge_0=nh,\quad n\in\mathbb N.
\end{equation}
The magnetic field of a magnetic point charge $g$ is given by
\begin{equation}\label{BMonop}
\vec B=\frac{g}{4\pi r^2}\vec e_r,
\end{equation}
with the singular density
\begin{equation}\label{DivBMonop}
\rho_M=\vec\nabla\vec B\ist{BMonop}g\,\delta(\vec r).
\end{equation}

\begin{figure}[h]
\centering
\includegraphics[scale=0.5]{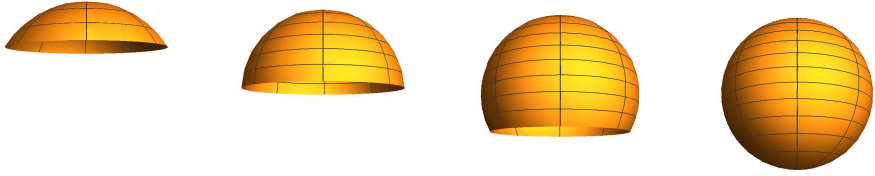}
\caption{Spherical caps $S(r,h)$ of height $h=r/2,\,r,\,3r/2$ and $2r$.}
\label{kalotten}
\end{figure}

The magnetic flux $\Phi$ through a spherical cap of height $h$~\footnote{This $h$ should not be confused with the Planck constant $h=2\pi\hbar$ in Eq.~(\ref{DiracQuantBed}).} of a sphere with radius $r$, see Fig.~\ref{kalotten} is given by
\begin{equation}\begin{aligned}\label{KalottenFluss}
\Phi&(r,h):=\int_{S(r,h)}\vec B\mathrm d\vec S
=r^2\int_0^{\vartheta_h}\sin\vartheta\,\mathrm d\mathrm\vartheta
\int_0^{2\pi}\mathrm d\varphi\,\vec e_r\vec B(r)=\\
&\ist{BMonop}r^2\int_0^{\vartheta_h}\sin\vartheta\,\mathrm d\mathrm\vartheta
\int_0^{2\pi}\mathrm d\varphi\,\frac{g}{4\pi r^2}=r^2(1-\cos\vartheta_h)\,2\pi\frac{g}{4\pi r^2}=\\
&=(1-\cos\vartheta_h)\frac{g}{2}=\frac{g}{2}\frac{h}{r}
\end{aligned}\end{equation}
and depends only on the ratio $h/r$.

The crucial step is the representation of the $\vec B$-field as curl of a gauge field $\vec A$
\begin{equation}\label{DivBMonop}
\vec B=\vec\nabla\times\vec A
\end{equation}
The most convenient gauge assumes rotational symmetry around an axis, e.g. the $z$-axis. It reads in polar coordinates $\vartheta$ and $\varphi$
\begin{equation}\label{EichungA}
\vec A=A_\varphi\vec e_\varphi.
\end{equation}
We can relate $A_\varphi$ to the magnetic flux
\begin{equation}\begin{aligned}\label{FlussB}
\Phi(r,h)&=\int_{S(r,h)}(\vec\nabla\times\vec A)\mathrm d\vec S
 =\int_{\partial S(r,h)}\vec A\mathrm d\vec s
 \ist{EichungA}r\sin\vartheta_h\int_0^{2\pi}\mathrm d\varphi A_\varphi=\\
 &=r\sin\vartheta_h2\pi A_\varphi(\vartheta_h)
\end{aligned}\end{equation}
and get the well-known Dirac potential~\cite{dirac:1931kp,dirac:1948um}
\begin{equation}\label{APhi}
A_\varphi(\vartheta)\iist{KalottenFluss}{FlussB}\frac{g}{4\pi r}\frac{1-\cos\vartheta}{\sin\vartheta}
 =\frac{g}{4\pi r}\tan\frac{\vartheta}{2}
 =\frac{g}{4\pi r}\frac{\sin\vartheta}{1+\cos\vartheta},\quad A_\vartheta=0.
\end{equation}
The magnetic flux through an area $S_{\mathcal C}$ bounded by a closed curve $\mathcal C$ is therefore given by
\begin{equation}\label{FlussS}
\Phi=\frac{g}{4\pi}\oint_{\mathcal C}\frac{1-\cos\vartheta}{r\sin\vartheta}
\vec e_\varphi\cdot\mathrm d\vec s.
\end{equation}
The Dirac potential has two types of singularities. A singularity at $r=0$ which is related to the point-like character of the monopole. The second singularity is the famous Dirac string, a singular line extending along the negative $z$-axis, see Eq.~(\ref{APhi}).

In the next section we show that choosing appropriate local coordinate systems on $\mathbb S^2$ we can get the same type of singularities for the connection field in the bundle of tangential spaces~\cite{Nakahara:2003aa}.

\section{Geometry of $\mathbb S^2$}\label{SecS2}
\begin{figure}[h]
  \centering
  \includegraphics[scale=0.5]{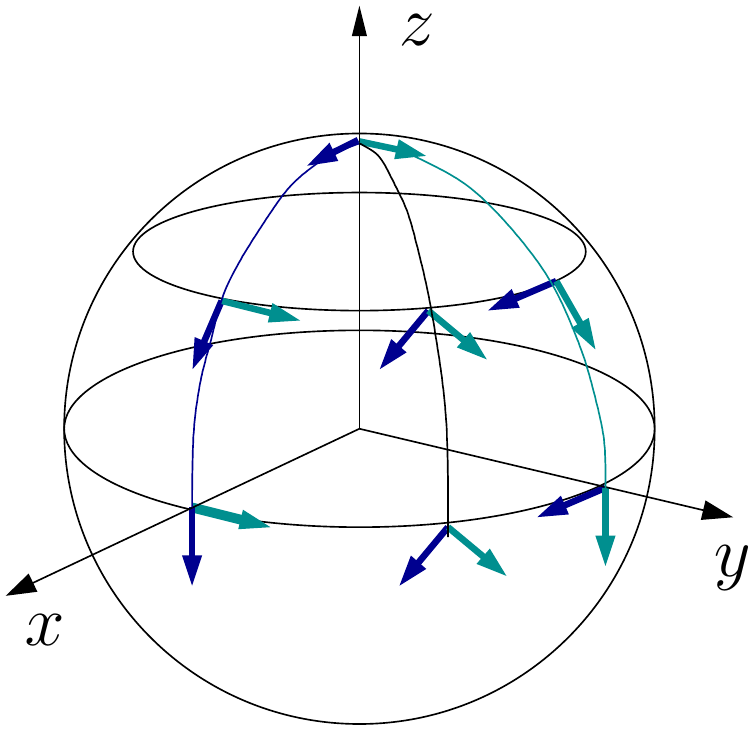}
  \caption{Local coordinate systems on $\mathbb S^2$, defined by parallel transport from the north pole along the meridians.}
  \label{dirackoord}
\end{figure}

\begin{figure}[h]
 \vspace{10mm}
 \centering
 \includegraphics[scale=0.6]{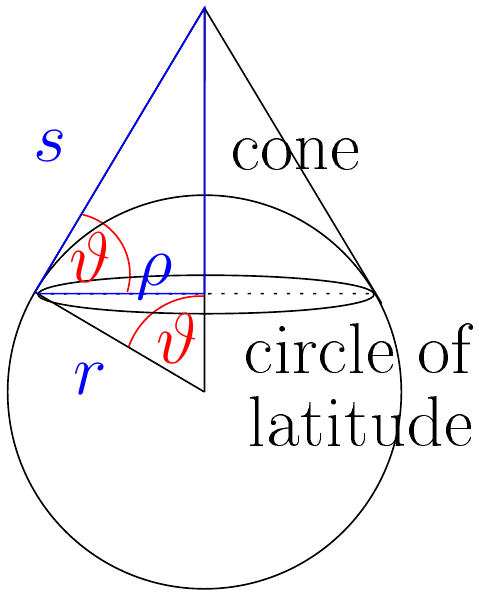}
 \includegraphics[scale=0.45]{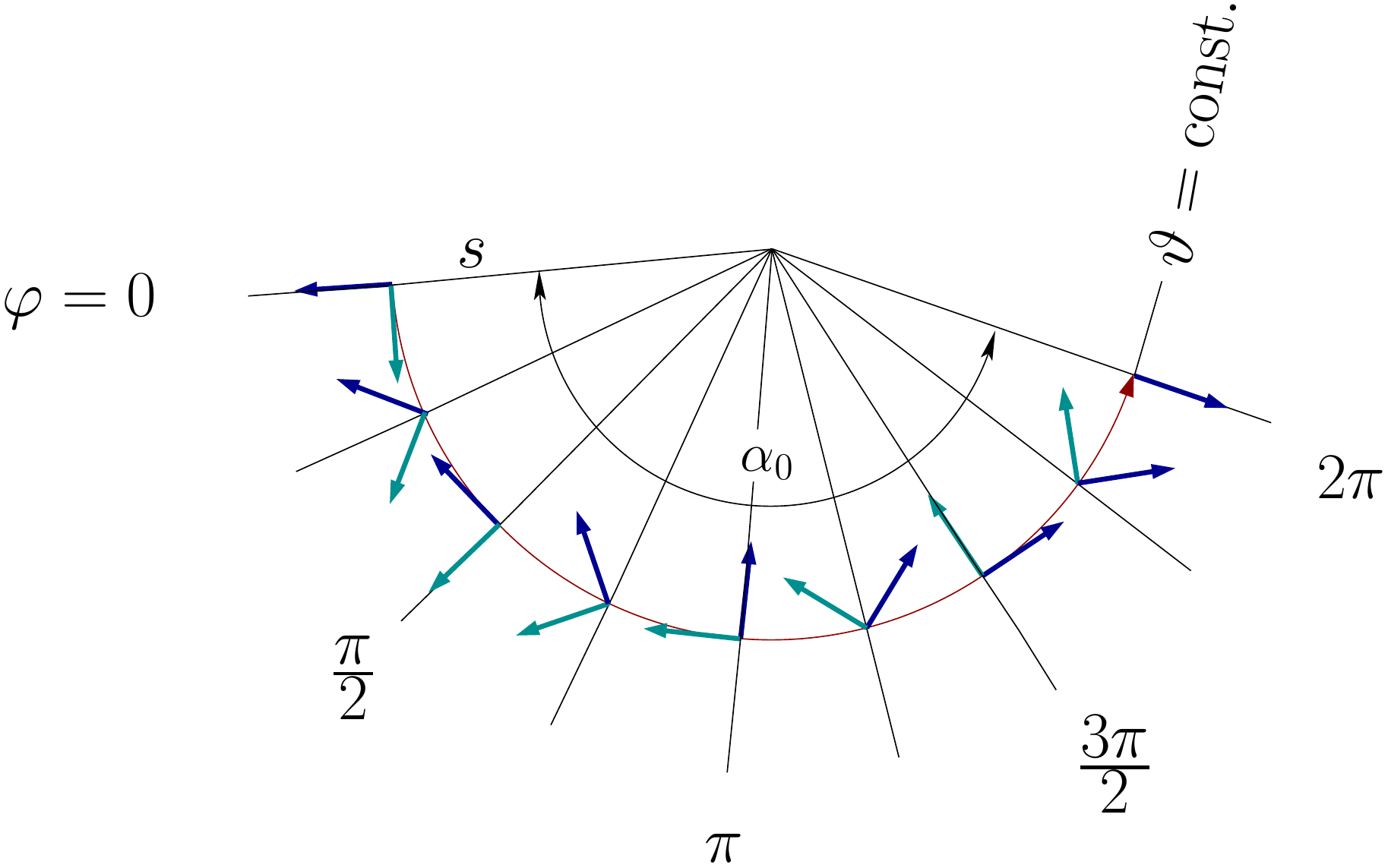}
 \caption{Left: Cone above a circle of latitude. Right: Map of the surface of such a cone to $\mathbb R^2$.}
 \label{kegel}
\end{figure}

In $\mathbb R^n$ the concept of parallel transport is well defined and applies also to cylindrical and conic surfaces~\cite{Aldrovandi:1995aa}. The neighbourhood of a meridian circle is such an infinitesimal small cylindrical surface. An orthonormal coordinate basis at the north pole, $\vartheta=0$, of $\mathbb S^2$ is transported along the meridians to every point of $\mathbb S^2$. Besides the south pole, $\vartheta=\pi$, these local bases are everywhere well defined, see Fig.~\ref{dirackoord}.
Moving along the surface from a point $\vec r$ along some infinitesimal line segment $\mathrm d\vec s=(r\mathrm d\vartheta,r\sin\vartheta\mathrm d\varphi)$ we find a rotation of the local $Zweibein$ by $\mathrm d\phi$. We describe it relative to the spherical basis $\vec e_\vartheta$ and $\vec e_\varphi$ by connection coefficients $\Gamma_\vartheta$ and $\Gamma_\varphi$
\begin{equation}\label{phaseFuerDphi}
\mathrm d\phi=\Gamma_\vartheta\mathrm d\vartheta+\Gamma_\varphi\mathrm d\varphi.
\end{equation}
According to the construction of the local bases we get
\begin{equation}\label{ZusTheta}
\Gamma_\vartheta=0.
\end{equation}
For $\Gamma_\varphi$ we construct cones above the circles of latitude, see the left diagram of Fig.~\ref{kegel}.
In the right diagram of Fig.~\ref{kegel} we map the surface of a cone to $\mathbb R^2$ and get a circular sector with arc length $2\pi\rho$ and central angle
\begin{equation}
\alpha_0=\frac{2\pi\rho}{s}\qquad\textrm{where}\qquad
\frac{\rho}{s}=\cos\vartheta.
\end{equation}
Moving with increasing $\varphi$ along a circle of latitude the local bases rotate in $\mathbb R^2$ in negative $\phi$-direction with the factor of proportionality $\Delta\phi/\Delta\varphi=(\alpha_0-2\pi)/2\pi=\cos\vartheta-1\le0$, see Fig.~\ref{kegel}
\begin{equation}\label{Gammaphi}
\Gamma_\varphi:=\frac{\Delta\phi}{\Delta\varphi}=\cos\vartheta-1
\quad\Rightarrow\quad\mathrm d\phi=\Gamma_\varphi\,\mathrm d\varphi.
\end{equation}

With the knowledge of the connection coefficients $\Gamma_\vartheta$ and $\Gamma_\varphi$ we can calculate the parallel transport along a closed curve $\mathcal C$ surrounding the surface $S_{\mathcal C}$~\footnote{For a closed curve, which has no self crossing, we have two areas, an area including the south pole and the complementary area $S_{\mathcal C}$.}. The rotational angle for a vector is identical to the area on the unit sphere. For the area on a sphere with radius $r$ we get
\begin{equation}\label{Kugelstueck}
S_{\mathcal C}=-r^2\oint_{\mathcal C}(\Gamma_\vartheta\mathrm d\vartheta
+\Gamma_\varphi\mathrm d\varphi)
\iist{ZusTheta}{Gammaphi}r^2\oint_{\mathcal C}(1-\cos\vartheta)\mathrm d\varphi
=r^2\oint_{\mathcal C}\frac{1-\cos\vartheta}{r\sin\vartheta}
\vec e_\varphi\cdot\mathrm d\vec s.
\end{equation}
The negative sign takes into account that we have to describe the direction of the parallel transported vector relative to the (passively) rotating basis, compare Fig.~\ref{kegel}. Up to the difference in the normalisation, $g$ for the magnetic flux and $4\pi r^2$ for the sphere of radius $r$, the two expressions~(\ref{FlussS}) and (\ref{Kugelstueck}) are identical.

\section{Conclusion and outlook}\label{SecSchluss}
A sphere has no singularity. Obviously the singularity in Eq.~(\ref{Kugelstueck}) has its origin in the inconvenient description of areas on $\mathbb S^2$ by tangential vectors as explained by the hairy ball theorem~\cite{Renteln:2013aa}, which states that there is no nonvanishing continuous tangent vector field on spheres of even dimension. If we describe $\mathbb S^2$ by radial vectors there is no singularity.

This raises the question of the relation of the gauge field~(\ref{FlussS}) to tangential vectors. This relation indicates that the gauge field describes the rotation of a local coordinate system on the $\mathbb S^2$ formed by a normalised three component vector field $\vec n$ of internal degrees of freedom. Rotations of the bases in the corresponding two-dimensional tangent spaces $\mathbb R^2$ are described by the group U(1), the gauge group of Maxwell's electrodynamics. A description of this internal $\mathbb S^2$ by these radial unit vectors $\vec n$ leads to a formulation of magnetic monopoles without Dirac string singularity. Such a description was given by Wu and Yang~\cite{wu:1975es}. In their description a vector field $\vec A_\mu$ is introduced as a vector-valued affine connection on this internal $\mathbb S^2$. By gauge transformations the direction of the $\vec n$-vectors can be aligned in 3-direction reducing the gauge degree of freedom to U(1) and the gauge field to the form of Eq.~(\ref{APhi}). The method to describe Dirac monopoles by a field of unit vectors $\vec n$ was also used in Ref.~\cite{deLemos:2009bn} to get rid of Dirac worldsheets in the Cho-Fadeev-Niemi representation of SU(2) Yang-Mills theory.

As Dirac mentioned there is no experimental encouragement for the existence of magnetic monopoles. But their most interesting property, the quantization of magnetic charge, can be converted to the quantization of electric charge by the above mentioned dual transformation, exchanging electric and magnetic fields. With this dual transformation it is possible to transfer the quantization of the electric charge into a classical formulation of electrodynamics. Moreover using the Wu-Yang formulation these electric monopoles do not need Dirac strings. Wu-Yang electric monopoles have only one type of singularity, the singularity at the origin characteristic for point-like charges. In Ref.~\cite{Faber:1999ia} a model of monopoles was suggested which removes also this second singularity. This model has quantized electric monopoles of finite mass. In Ref.~\cite{Faber:2002nw} the dynamics of this model was compared to Maxwell's electrodynamics.






\bibliographystyle{SciPost_bibstyle.bst}
\bibliography{literatur}





\end{document}